# Integrated ridesharing services with chance-constrained dynamic pricing and demand learning

Tai-Yu Ma*, Sylvain Klein

Luxembourg Institute of Socio-Economic Research (LISER), 11 Porte des Sciences, L-4366 Esch-sur-Alzette, Luxembourg

* Corresponding author

## Abstract

The design of integrated mobility-on-demand services requires jointly considering the interactions between traveler choice behavior and operators' operation policies to design a financially sustainable pricing scheme. However, most existing studies focus on the supply side perspective, disregarding the impact of customer choice behavior in the presence of co-existing transport networks. We propose a modeling framework for dynamic integrated mobility-on-demand service operation policy evaluation with two service options: door-to-door rideshare and rideshare with transit transfer. A new constrained dynamic pricing model is proposed to maximize operator profit, taking into account the correlated structure of different modes of transport. User willingness to pay is considered as a stochastic constraint, resulting in a more realistic ticket price setting while maximizing operator profit. Unlike most studies, which assume that travel demand is known, we propose a demand learning process to calibrate customer demand over time based on customers' historical purchase data. We evaluate the proposed methodology through simulations under different scenarios on a test network by considering the interactions of supply and demand in a multimodal market. Different scenarios in terms of customer arrival intensity, vehicle capacity, and the variance of user willingness to pay are tested. Results suggest that the proposed chance-constrained assortment price optimization model allows increasing operator profit while keeping the proposed ticket prices acceptable.

## 1. Introduction

Mobility on demand (MOD) services such as taxis, paratransit, ride-hailing, and microtransit, etc. have received increasing interest as they are user-centered, convenient, and present huge potential to alleviate the inconvenience of conventional fixed-route transit (Kwoka-Coleman, 2017). With the emerging autonomous vehicle (AV) technology and the tendency for Mobility-as-a-Service (MaaS), future urban mobility will be shaped towards a more automated, clean, and seamless multimodal transportation service. In line with this research challenge, in recent years there have been several initiatives based on public–private partnerships to provide transit-integrated mobility services that extend a classical monomodal MOD service to an integrated multimodal service (Etherington, 2017). The integrated MOD services have great potential for reducing operating costs and customers' waiting times and increase the service capacity and ridership of transit systems (Ma et al., 2019a). Like many existing microtransit services, a transport network company (TNC) operates a fleet of capacitated vehicles with flexible routes to provide customers with either direct rides or combined rides (e.g., rideshare+transit) by integrating rideshare as a part of the multimodal trip for customers. An example is illustrated in Fig. 1. Customers book their trips in advance via smartphone or internet. The operator proposes a service assortment with relevant attributes, i.e., fare, waiting time, in-vehicle riding time, etc. Customers make their choice based on their implicit travel preferences in a multimodal market. The integrated MOD services are particularly suitable for rural areas due to low accessibility to public transport and scattered travel demand. Due to the uncertainty of customer demand and the competition from other transport means, most (integrated) MOD services adopt a low fixed-fare strategy supported by subsidies from the government (Go Centennial Final Report, 2017). Several failed microtransit services such as Helsinki's Kutsuplus (Sulopuisto, 2016) and San Francisco's Chariot (Hawkins, 2019) show the importance of travel demand estimation and pricing strategy for the successful deployment of MOD services. However, most studies mainly focus on vehicle dispatching and routing policy design (Agatz et al., 2012) or pricing (Atasoy et al., 2015; Sayarshad and Chow, 2015) without jointly modeling customer mode choice behavior in a competing multimodal market. Moreover, when considering customer choice behavior, most studies require conducting a relevant mobility survey to calibrate the model parameters (Wen et al., 2018).

Recent efforts have been devoted to the development of agent-based simulation tools for the evaluation of MOD services. For example, Levin et al. (2017) proposed an agent-based simulation framework by considering a realistic traffic flow model for dynamic ridesharing impact evaluation using shared autonomous vehicles (SAV). Customer mode choice behavior is not considered in this study. Chen et al. (2019) propose an agent-based model for dynamic ridesharing simulation evaluation in a multimodal market. Heuristic rules are proposed to model agent mode choice decisions by considering maximum waiting time and travel time for three mode alternatives, i.e., drive-alone, ridesharing, and public transit. Hörl et al. (2019) proposed an agent-based simulation framework using MATSim to evaluate the impact of deploying an automated taxi fleet on Paris. A discrete choice model is applied to simulate user mode choice decisions in a multimodal market. Their analysis shows the impact of SAV is significantly influenced by the interplay of operational cost, demand, and ticket prices. Vosooghi et al. (2019) conducted a comprehensive simulation evaluation on the impact of SAV using MATSim by considering all operational and demand-side travel preference variations. The authors emphasize the importance of considering dynamic demand in a multimodal market for the impact evaluation of SAV services. Ma et al. (2018, 2019a) developed a dynamic bimodal ridesharing model, allowing the operator to provide a suite of options to minimize operation costs. Their simulation case study of New York City shows that integrating rideshare services with public transport could reduce operation costs and increase public transport ridership. However, demand is assumed to be fixed, and customer preference and pricing issues were not considered. It is desirable to jointly consider the interactions of traveler choice behavior and operator operation and pricing policy to evaluate ridership and achieve the desired objective (e.g., social welfare/profit maximization). Some recent research has started looking at user–operator assignments based on the assignment game approach in a static setting (Rasulkhani and Chow, 2019; Ma et al., 2019b). In contrast to past studies, we focus on dynamic pricing with demand learning for an integrated microtransit service considering the interaction of supply and demand in a multimodal market.

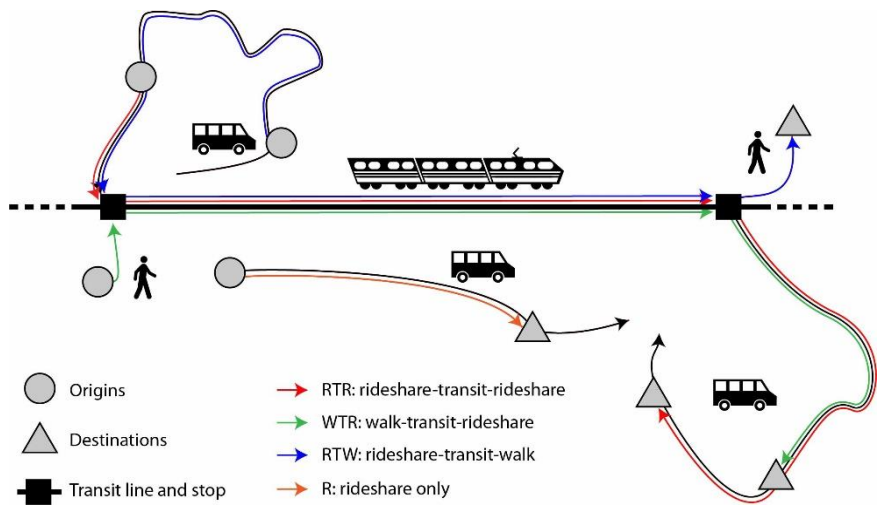

Fig. 1. Integrated microtransit system. Based on request origins and destinations, an operator proposes two options: "rideshare only I" and "rideshare+transit (RT)", with differentiated ticket prices and levels

of service (out-of-vehicle time, in-vehicle time). The latter includes rideshare–transit–rideshare (RTR) and walk–transit–rideshare (WTR/RTW) or vice versa.

To the best of our knowledge, none of the existing integrated MOD operation models takes into account the supply–demand interactions for dynamic price optimization. Unlike existing surge pricing methodology, which is based on friction of supply and demand, our approach integrates customer's preference into dynamic pricing settings under uncertainty to maximize the operator's profit. This is the first study to demonstrate an integrated microtransit operation policy design with demand learning and constrained dynamic pricing in a competing transportation market.

The contributions of this paper are summarized as follows.

– We propose a modeling framework for a dynamic integrated microtransit system with customer demand learning and dynamic pricing. The proposed model learns customer mode choice preferences in a multimodal market based on the operator's historical ride data. We demonstrate that the choice model can be calibrated and updated regularly.
– We propose a dynamic pricing model based on chance-constrained assortment optimization. Unlike existing methodologies (Atasoy et al., 2015; Qiu 2017), we consider customers' willingness-to-pay to constrain proposed ticket prices. A nested logit model (NL) is proposed to capture the correlation of different choice alternatives in the assortment price optimization.
– We conduct a series of simulation studies on a test network to evaluate the impact of the proposed methodology on an operator's revenue, operating cost, and customer inconvenience. Several scenarios related to different customer arrival intensities, vehicle capacity, and the variance of customers' willingness to pay are tested. The results show that the proposed chance-constrained pricing model would result in reasonable ticket prices and significantly increase operator profit, with higher "rideshare+transit" ridership shifted from "rideshare only".

This paper is organized as follows. In Section 2, we review existing studies on integrated MOD service planning, dynamic pricing methodology, and customer demand learning. In Section 3, we propose a modeling framework for an integrated microtransit service with dynamic pricing. Section 4 presents a chance-constrained dynamic assortment pricing optimization model for the integrated MOD services. In Section 5, numerical studies are conducted to evaluate the performance of the proposed model under different demand intensities. We also present a sensitivity analysis to evaluate the impact of key decision parameters. Finally, conclusions are drawn and future extensions are discussed.

## 2. Literature review

Existing integrated rideshare service studies mainly focus on vehicle dispatching and routing policy to minimize operating costs and customer inconvenience (Häll, 2006; Ma et al., 2019a, among many others). Travel demand is assumed to be fixed, without considering the co-existence of other transport means. The traditional traffic assignment methodology is not suitable for modeling user–operator traffic flow for MOD services (Chiu et al., 2011; Lebacque et al., 2007) as demand for a rideshare service depends on its system characteristics alone, without considering the competition from other transport means and customer preference. Some recent studies have started looking at these issues (Liu et al., 2018; Wen et al., 2018; Hörl et al., 2019; Vosooghi et al., 2019). These studies show travel demand is sensitive to service pricing as well as its level of service. However, how to integrate customer travel choice preference into the MOD service pricing remains an issue for a successful deployment of emerging MOD services.

Dynamic pricing is a price-based revenue management strategy in which sellers adjust the selling prices of goods over time based on customer demand to maximize the expected profit of sales. It is a widely adopted revenue management strategy in business or industrial sectors such as retailers, electricity, air transportation, and ride-sourcing markets (den Boer, 2015; Talluri and Van Ryzin, 2006). In recent years, dynamic pricing has been receiving much research interest in the ridesharing and ride-hailing market. Based on whether there exists an intermediary platform to match riders and drivers, we can distinguish the market as a one-sided or two-sided market. For the one-sided market, Sayarshad and Chow (2015) propose a non-myopic vehicle dispatching and pricing model for the dial-a-ride problem based on an M/M/1 queuing system. Pricing is determined to achieve a social optimum. Sayarshad and Gao (2018) further extend this study based on a multi-server queuing system and propose socially efficient and profit-maximization pricing schemes. Their case study using New York City taxicab data shows that the proposed model could significantly increase social welfare compared to the model based on the single queuing system. In line with the queuing-theory-based approach, Chen and Wang (2018) propose a pricing model to maximize the social welfare of a last-mile transportation service based on a multi-service queue system. Chen and Kockelman (2016) propose an agent-based modeling framework to evaluate the potential impact of different pricing schemes for deploying shared autonomous electric vehicles. Three pricing schemes, i.e., distance-based, origin-/destination-based, and mixed pricing schemes, are tested under the multinomial logit model (MNL) model to model traveler mode choice behavior by considering private-use AVs, shared-use AVs, and transit. For the two-sided market, Banerjee et al. (2015) developed a queuing-system-based dynamic pricing model and evaluate the impact on the revenue and the number of matches of the platform. Yan et al. (2019) survey the current matching and dynamic pricing mechanisms of ride-hailing. They find that surge pricing, i.e., dynamically adjusted ticket prices during high-demand periods, would reduce customer waiting times and increase service reliability and capacity.

While these studies propose different dynamic pricing schemes to adjust supply–demand mismatch, the traveler decision process in a multimodal market is not considered. Several recent studies integrate demand modeling in dynamic pricing based on the assortment optimization approach. Atasoy et al. (2015) propose an MNL-based assortment optimization model to maximize an operator's expected profit for three types of services: taxi, shared-taxi, and mini-bus services, given relevant constraints on vehicle operations (i.e., scheduling and/or capacity). Qiu (2017) integrates customer choice behavior based on the MNL model for the assortment price optimization of MOD services. A reference mode is assumed when customers do not choose to use the proposed MOD services. From a customer's point of view, it is relevant to consider mode choice behavior in a multimodal market. Moreover, most assortment-optimization-based approaches fail to consider customer willingness to pay, resulting in an unreasonably high ticket price setting.

Traditionally, it is desirable to conduct expensive revealed/stated preference surveys to collect travel preference data for demand function estimation. However, it is possible to learn the characteristics of customer demand via historical sales data, without additional surveys. For example, Talluri and Van Ryzin (2006) propose an estimation procedure based on the expectation-maximization method for MNL-based choice model estimation using historical purchase data. Sauré and Zeevi (2013) propose an assortment-optimization model with demand learning from customer historical data to estimate an MNL-based choice model. Bertsimas and Perakis (2006) assume that unknown demand is a function of price, which follows some probability distribution family. The authors propose a pricing mechanism based on dynamic programming to jointly learn demand parameters and set up optimal prices. The reader is referred to den Boer (2015) for a comprehensive literature review of dynamic pricing with demand learning. Table 1 summarizes the recent literature on dynamic pricing for MOD operation policy design.

Table 1. Summary of recent literature on dynamic pricing for on-demand mobility services

| Reference | Service type | Characteristics | Mode choice modeling and demand learning |
|---|---|---|---|
| Qiu (2017); Atasoy et al. (2015) | Mobility on-demand services | Logit-model-based assortment optimization | No |
| Sayarshad and Chow (2015); Sayarshad and Gao (2018) | Dynamic dial-a-ride and pricing problems | Non-myopic pricing based on service queue approximations | No |

| Zha et al. (2018) | Ride-sourcing | Surge pricing | No |
| Banerjee et al. (2015) | Ridesharing in a two-sided market | Surge pricing based on supply–demand imbalance using a queuing model | No |
| Wang et al. (2016); Zhang and Ukkusuri (2016) | Taxi | Market equilibrium model for a two-sided market | No |
| Chen and Wang (2018) | Last-mile transportation service | Constrained optimization under service-level constraints | No |
| Chen and Kockelman (2016) | Shared autonomous electric vehicle fleet management and pricing | Distance-/origin-/destination-based dynamic pricing schemes | Yes |
| Karamanis et al. (2018) | Ride-sourcing using autonomous vehicles | One-sided market, utility-based dynamic pricing considering travel time, waiting time, and prices for mode choice decision making. | Nested logit model for modeling choice between private, shared rides, and transit |

## 3. Modeling dynamic microtransit with transit transfer

We consider a dynamic microtransit (ridesharing) service planning problem with dynamic pricing in a multimodal transportation market. A microtransit service provider operates unimodal (rideshare only) and bimodal (rideshare+transit) ridesharing services based on a fleet of homogeneous capacitated vehicles $V = \{v_1, v_2, \ldots, v_{|V|}\}$. The bimodal ridesharing service is a service integrated with the existing transit network to provide a seamless multimodal solution for passenger transportation. We model the problem on a direct graph $G(N, E)$, where $N$ is a set of nodes and $E$ is a set of links. The set of nodes includes customer pick-up and drop-off locations, vehicle depots, and transit stations. Travel time $t_{ij}$ from node $i$ to node $j$ is the average travel time in a studied area. We assume that demand is unknown, and customer arrivals are stochastic following a Poisson process with an arrival intensity of $\lambda$. Empirical customer arrival patterns can be applied for realistic applications. Customer mode choice preferences are assumed unknown and considered in light of the presence of other transport means. We consider a day-to-day mode choice setting under stochastic demand. Traveler choice decisions are based on random utility theory in a multimodal market (Train, 2009). As different mode choice alternatives might be correlated, the nested logit model is applied to capture its correlation.

The characteristics of the system are as follows.

− A fleet of capacitated vehicles are operated by a dispatching center that determines vehicle routing, pricing, and dispatching. All vehicles are initiated at initial depots at the beginning of each day and return to the depots at the end of that day. When a new customer needs a ride, he/she sends a ride request to the operator from his/her smartphone. The request includes relevant information comprising the desired pickup and drop-off locations and pick-up time.
− Upon receiving a request, the operator first identifies available vehicles within a given service range (e.g., 5 km service radius of customer pickup locations) as potential candidates for vehicle dispatch. The service-range-based dispatching policy has been widely adopted in realistic applications for ride-hailing services. Based on the designed routing policy (Eq. (1) and (2)) and vehicle capacity constraints, the operator proposes two options: rideshare only (R) or rideshare+transit (RT). A quote is then sent to the customer with relevant information including fare, walking time, waiting time, in-vehicle travel time, and transfer time whenever relevant. If the new customer accepts the offer, the dispatching center assigns a vehicle to serve the customer and updates the system states. Otherwise, the customer rejects the offer and chooses an alternative among available transport means. With the aim of improving the MOD service, the new customer is asked to report their mode choice in all cases (to use or not use the MOD services). We also assume that once the customer's decision is made, he/she will carry out his/her decision.

– For the idle vehicle repositioning strategy, we apply the myopic idle vehicle relocation policy, known for its computational efficiency (Ma et al., 2019a; Sayarshad and Chow, 2017). The idle vehicle relocation policy is activated at the beginning of each rebalancing epoch (e.g., 10–15 minutes). An en-route switching policy is applied to allow in-transition idle vehicles to pick up new customers.

We apply a non-myopic vehicle dispatching and routing policy to anticipate future system state (Sayarshad and Chow, 2015). The non-myopic policy assigns a new request to vehicles with the minimum additional insertion cost considering future delay costs, as follows.

$$\{v^*, x_t^{v^*}\} = \text{argmin}_{v,x}[c\,(v, x_t^v) - c(v, x\,)] \qquad (1)$$

where $x_t^v$ is a new tour after a new request is inserted. $c(v, x)$ is a cost function for vehicle $v$ operating tour $x$, defined in Eq. (2) as

$$c(v, x) = \gamma T(v, x) + (1 - \gamma)[\beta\,T(v, x)^2 + \sum_{n \in P_v} Y_n(v\,, x)] \qquad (2)$$

where $T(v, x)$ is the length (measured in time) of tour $x$. $Y_n(v\,, x)$ is the journey time, including waiting time and in-vehicle travel time for passenger $n$. $P_v$ is the set of passengers on board of vehicle $v$. Eq. (2) considers the weighted sum of the system cost and on-board customer inconvenience based on a user-defined parameter $\gamma$. Note that the above vehicle dispatching policy is a post-optimization method, which minimizes additional system costs by anticipating future system delays when inserting a new request. The practitioner can further consider time windows constraints associated with customer's pickup and drop-off locations and apply dedicated algorithms for vehicle routing and dispatching (Häme, 2011).

When the operator receives a request, two options (i.e., rideshare only and rideshare+transit) are considered for each nearby vehicle within a pre-defined search radius. The generation of the two options is described as follows.

-Rideshare only: Given a set of nearby vehicles within the service range of a new request, a least-cost vehicle dispatch with respect to Eq. (1) is calculated. We use insertion-based heuristics (Mosheiov, 1994) to solve the underlying travelling salesman problem with pick-up and delivery (TSPPD). The reader is referred to Sayarshad and Chow (2015) for more details.

- Rideshare+Transit (RT): This option divides a trip into a sequence of trip legs. It considers a "rideshare" leg from a user's origin to a "least-time" (in terms of door-to-door trip, described in Algorithm 1) transit station and a "transit" leg to the final destination. The egress part from an exit station to the customer's destination is assumed to occur on foot. Note that other combinations (i.e., rideshare–transit–rideshare and walk–transit–rideshare) can be searched for realistic applications (Ma et al., 2019a).

The modeling framework of the integrated MOD system with dynamic pricing and demand learning is depicted in Fig. 2. The details of vehicle dispatching are described in Algorithm 1.

**Algorithm 1.** Dynamic integrated microtransit with price optimization

1. Initialization. Set day=1. Run the simulation without dynamic price optimization until all customers served. Initial parameters in the demand function.
2. Day:= Day +1. Randomly generate customer arrivals. Upon arrival of a new customer r, update vehicles' current state from the time of previous request r-1

3. For each vehicle within the service range of customer r,
    i. Compute the following two options given vehicle's capacity constraints:
       – Rideshare only (R): Solve the TSPPD problem based on the re-optimization-based insertion algorithm (Mosheiov, 1994) to obtain a potential tour that minimizes the generalized cost function of Eq. (1).
       – Rideshare+Transit (RT): Two steps are applied. First, identify entry and exit stations based on minimizing total door-to-door travel time over k nearest entry and k nearest exit stations related to the origin and destination of r. We calculate travel time from origin to the entry station using rideshare by solving the TSPPD problem. Travel time in the transit system includes waiting time, transfer time, and in-vehicle travel time.
   ii. Determine optimal fare adjustments for R and RT options based on the chance-constrained assortment pricing optimization (described in Section 4).
4. Send the quote to customer r. Customer makes a mode choice decision accordingly. Customer reports his/her choice decision via the operator's app. If the RT option is chosen, update customer's drop-off location to the corresponding station of the proposed RT option.
5. Update the tour of the assigned vehicle.
6. Continue until all customers are served.
7. Update the mode choice model estimation based on the collected historical data. Go to Step 2.

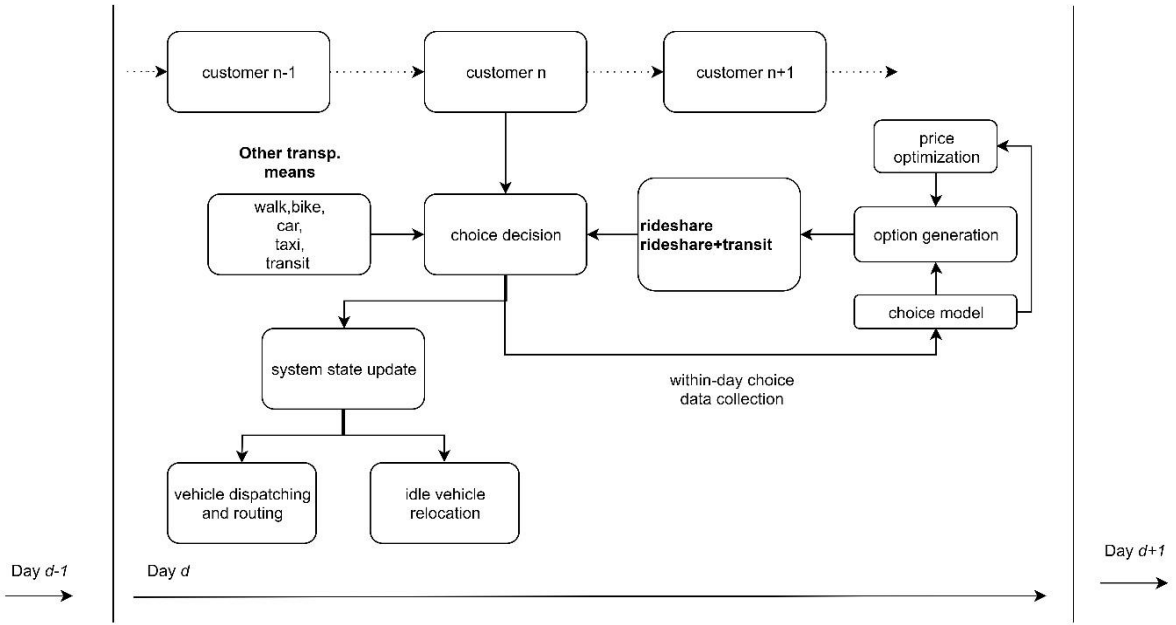

Fig. 2. Modeling framework of integrated microtransit with dynamic pricing and demand learning.

Fig. 2 shows the proposed modeling approach, which jointly considers the interaction of supply and demand in a multimodal market. The decision to use the MOD service (or not) depends on the value of utility of different choice alternatives. Note that for simplification, we assume that the considered population (who is interested in using the system) is a small part of the total population; an individual's mode choice decision has a marginal impact on the congestion of the existing transport infrastructure. One can relax this assumption by implementing more realistic transportation system characteristics to capture the congestion effect due to the collective behavior of each individual's travel decision.

**4. Dynamic pricing with customer mode choice preference learning**

Unlike most MNL-based assortment price optimizations, we model customer mode choice behavior based on the NL model to capture the correlation between different mode choice alternatives (Train, 2009). Let $C$ denote the universal mode choice set in a multimodal transportation market. We assume that the dependency among different choice alternatives is structured into two levels as shown in Fig. 3. Let $C_1$ be the set of nests, defined as $C_1 = \{\text{nonmotorized}, \text{auto}, \text{public transport (PT)}\}$. We assume that the IIA (independence of irrelevant alternatives) property holds among the nests. Let $U_{nj}$ be the utility obtained by a customer $n$ when choosing alternative $j$ (Train, 2009):

$$U_{nj} = V_{nj} + \varepsilon_{nj}, \forall j \in J_m, m \in C_1 \tag{3}$$

where $V_{nj}$ is the determinant part of the utility, which depends on alternative-specific and individual-specific attributes. $\varepsilon_j$ is a random variable following a Gumbel distribution capturing unobserved influence factors. Under the linear utility assumption, we specify $V_{nj}$ as

$$V_{nj} = \beta_{ASC} + \sum_k \beta_k X_{nk} + \sum_j \beta_j X_{nj} \tag{4}$$

where $\beta_{ASC}$ is an alternative-specific constant and $\beta_k$ and $\beta_j$ are individual- and alternative-specific parameters, respectively. $X_{nk}$ and $X_{nj}$ are a set of individual-specific and alternative-specific attributes, respectively. For simplicity, we consider only alternative-specific attributes, including out-vehicle travel time (OVTT, including walking time, waiting time, and transfer time), in-vehicle travel time (IVTT), and fare (monetary cost), i.e., $X_{nj} = \{OVTT_j, IVTT_j, f_j\}, \forall j \in C$. Based on the NL model, the probability of customer $n$ choosing alternative $j$ can be written as

$$P_{nj} = P(j|m)P(m), \forall j \in J_m, \forall m \in C_1 \tag{5}$$

where $P_n(j|m) = \frac{e^{V_{nj}/\mu_m}}{\sum_{j \in J_m} e^{V_{nj}/\mu_m}}$ and $P_n(m) = \frac{e^{\mu_m I_m}}{\sum_{s \in C_1} e^{\mu_s I_s}}$.

The term $P(m)$ is the choice probability of nest $m$, and $P(j|m)$ is the conditional probability of choosing $j$ within nest $m$. $I_m$ is the inclusive value related to nest $m$, which measures the aggregate attractiveness of that nest, defined as

$$I_m = \ln\left(\sum_{j \in B_m} e^{\frac{1}{\mu_m}V_{nj}}\right) \tag{6}$$

where $\mu_m > 0, \forall m \in C_1$, is the NL model parameters. $B_m$ is the subset of choice alternatives within nest $m$. $1-\mu_m$ represents the degree of correlations of unobserved utility among alternatives in a nest.

Given a set of individuals, $n = 1, \ldots, N$, the log-likelihood function can be written as

$$LL(\beta) = \sum_n \sum_{j \in C} y_{nj} \ln P_{nj} \tag{7}$$

where $y_{nj}$ is an indicator taking a value of 1 if customer $i$ chooses alternative $j$. $P_{nj}$ is the choice probability defined by Eq. (4). As $P_{nj}$ takes a closed form, the maximum-likelihood approach can be applied for the parameter estimation without difficulty.

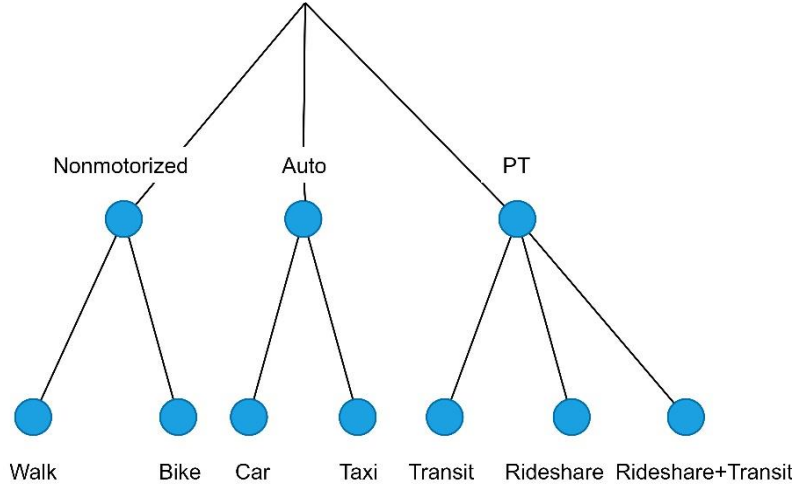

Fig. 3. Nested structure of choice options for a traveler's mode choice decision

We consider a myopic assortment pricing optimization problem to maximize the operator's expected profit for the current request (Atasoy et al., 2017). Optimizing the ticket price for a given time horizon requires applying an approximate dynamic programming approach to estimate the future system state; this is a well-known and difficult issue for the DARP problem (Sayarshad and Chow, 2015; Qiu, 2017). We limit ourselves to optimizing ticket prices for each single request to maximize the expected profit of the operator. According to the information provided with each new customer, i.e., desired origin, destination, desired pickup time, and choice decision, the operator constitutes the alternative-specific attributes for each alternative over time. We can then estimate the NL-based choice model for assortment price optimization to maximize the expected profit of the operator.

The profit of MOD option $j \in J_O$ with price adjustment $\delta_j$ is defined as

$$f_j - c_j + \delta_j, j \in J_O \tag{8}$$

where $f_j$ is the fare for option $j$ before price optimization. $c_j$ is the operation cost of option $j$. $\delta_j$ is the decision variable of the fare adjustment for option $j$. $J_O$ is the set of options provided by the operator, i.e., $J_O = \{\text{Rideshare}, \text{Rideshare} + \text{Transit}\}$. The dynamic pricing problem consists in determining optimal ticket prices for the assortment $J_O$ to maximize the expected profit of the current ride request. Under the existing unconstrained assortment price optimization approach (Atasoy et al., 2015; Qiu, 2017), the optimal prices might be too high, which is not desirable. To overcome this issue, we consider customers' willingness to pay as constraints and propose a chance-constrained assortment price optimization model. In doing so, the optimal ticket prices will be bounded at the customer's willingness to pay for that trip. Following Breidert (2005), we consider **a customer's willingness to pay $\varpi$ as his/her maximum price to use the MOD service** (i.e. rideshare and rideshare + transit). We assume $\varpi$ is a random variable, defined as Eq. (9).

$$\varpi = w + \varepsilon \tag{9}$$

where $w$ is the perceived cost of the reference product (i.e., travel cost of using car for that trip, assuming as a linear function of travel time of trip) and $\varepsilon$ is the differentiation value following a hypothetical normal distribution with mean 0 and variance $\sigma = s^2$. Given that customer willingness to pay is a random variable, the assortment price optimization becomes a chance-constrained optimization problem as in Eq. (10)–(11).

$$max \sum_{j \in J_O} P_{nj}(\delta_j)(f_j - c_j + \delta_j) \tag{10}$$

s.t.

$$f_j + \delta_j \leq \varpi_n, \quad \forall j \in J_O \tag{11}$$

The objective function (10) is to find optimal fare adjustments to maximize the expected profit of the current ride request. $P_{nj}(\delta_j)$ is the choice probability for customer $n$ choosing MOD option $j \in J_O$ after the price adjustment $\delta_j$. The price adjustment $\delta_j$ changes customer's utility perception $V_{nj}(\delta_j)$ given the fare adjustment $\delta_j$, defined as

$$V_{nj}(\delta_j) = V_{nj} + \beta_f \delta_j \tag{12}$$

where $\beta_f$ is the coefficient of fare in the NL model. We then apply Eq. (12) into Eqs. (5)-(6) to calculate $P_{nj}(\delta_j)$ in the objective function.

Equation (11) bounds ticket prices on customer willingness to pay $\varpi_n$, which is now stochastic constraints because of the presence of a stochastic term $\varpi_n$. By introducing Eq. (9) in (11), Eq. (11) becomes Eq. (13).

$$f_j + \delta_j - w \leq \varepsilon, \quad \forall j \in J_O \tag{13}$$

Considering the concept of $\alpha$-reliability (e.g., if $\alpha = 0.05$, there is a 95% chance that the optimal ticket price will not exceed the customer's willingness to pay), then Eq. (13) can be expressed deterministically as a chance constraint (14).

$$\Phi[f_j + \delta_j - w] \leq 1 - \alpha, \quad \forall j \in J_O, \tag{14}$$

where $\Phi(z) = \Pr(Z \leq z)$ is the cumulative density function of Z.

According to Shapiro et al. (2009), the nonlinear constraint Eq. (14) can be transformed into a linear inequality as Eq. (15):

$$f_j + \delta_j - w \leq Z_{1-\alpha} s. \tag{15}$$

Eq. (15) means that the higher variation of customer's willingness to pay is, the higher ticket price could be set up to maximize the operator's profit. In reality, **an operator needs to further calibrate the variation of user's willingness to pay to better capture customer's sensitivity to ticket price**.

The transformed linear constrained optimization problem can be solved without difficulty using the classical Lagrange multiplier methods (Bertsekas, 2016). We use the commercial Matlab solver fmincon to obtain the solution efficiently.

## 5. Numerical experiments

In this section, we apply the model on numerical examples to quantify the effect of chance constrained dynamic pricing on the operator's profit and user's inconvenience. Our goal is to demonstrate the methodology via numerical simulations. The applications on real-world services would be future extensions of this work by considering time-window constraints, synchronization at transit stations to minimize customer's transfer inconvenience, and other realistic considerations. The numerical experiment is divided into two parts. First, we discuss the generation of the test instance and the utility function specification. Then we analyze the computational results for different arrival intensities. Finally, a sensitivity analysis is conducted to evaluate the impact of different decision parameters.

## 5.1. Data generation and utility function specification

We consider a 20 km × 20 km region with a transit network. Demand is assumed to be randomly distributed following a Poisson distribution with the arrival intensity of $\lambda$ (customers/hour). The spatial distribution of customer pick-up and drop-off points are shown in Fig. 4. The entire region is divided into a 16 zones with an identical size of 5 km × 5 km. A number of vehicles are initially deployed at the centroid of each zone by a microtransit operator. We assume vehicles need to return to the depots after service. For the transit network, six transit lines (i.e., North–South, East–West, Northeast–Southwest, Northwest–Southeast, two ring-lines) operate in both directions with pre-defined timetables. The capacity and frequency of transit lines are assumed to be sufficient to absorb demand without congestion at the transit stations.

As we have no empirical stated/revealed survey data to calibrate demand function of users, we will follow the state of the art random utility theory based on hypothetical utility function specification (Train, 2009). A set of mode choice alternatives are assumed to be available for each customer including: walk, bike, car, taxi, transit, rideshare, rideshare+transit for their mode choice decision. We assume that these alternatives have an implicit nested structure as shown in Fig. 3. The deterministic part of the customer utility function is specified as a function of alternative-specific attributes as Eq. (16), considering main mode choice influence factors (Wen et al., 2018; Hörl, et al,2019):

$$V_{nj} = ASC_j X_{nj} + \beta_{OVTT} OVTT_{nj} + \beta_{IVTT} IVTT_{nj} + \beta_c C_{nj}, \tag{16}$$

where
$IVTT_{nj}$ is the in-vehicle travel time of mode $j$ for customer $n$;
$OVTT_{nj}$ is the out-of-vehicle travel time of mode $j$ for customer $n$;
$C_{nj}$ is the travel cost of mode $j$ for customer $n$;
$X_{nj}$ is a mode-specific dummy variable for mode $j$ for customer $n$;
$ASC_j$ is alternative-specific constant for mode $j$.

The parameters of the NL models are assumed to be unknown for the operator. Via the collection of the customer choice data over time, the operator can calibrate the parameters of the choice model over time (e.g., at the end of each day). The unknown true coefficients are assumed to be $\beta_{OVTT} = -0.032$, $\beta_{IVTT} = -0.023$ and $\beta_c = -0.074$, adapted from the case study of Liu et al. (2018). For simplicity, we assume the alternative-specific constant (ASC) value to be 0 for all alternatives. The unknown true scaling parameters are assumed to be $(\mu_{non-motorized}, \mu_{auto}, \mu_{pt}) = (1, 2, 2)$. Note that one can calibrate the ASCs based on the observed market share of realistic applications (Train, 2009).

The experiment setting related to the characteristics of different transport modes is as follows.

- Walk: The average speed is assumed to be 5 km/ hour.
- Bike: The average speed is assumed to be 16 km/hour.
- Car: The average speed is assumed to be 25 km/hour. The average cost of driving by car is assumed to be 0.33 $/km (Victoria Transport Policy Institute, 2009). No parking cost is considered.
- Taxi: An average 5 minute waiting time is assumed. The taxi fare is $3.0 (initial cost) + $1.56 per kilometer traveled.[1]
- Transit: Travel time includes access times, in-vehicle travel time, and waiting times at stations. The access mode from and to the nearest stations is assumed to be on foot. Transit speed is assumed to be 60 km/hour. An average waiting time of 5 minutes is assumed, given the headway of 10 minutes. The transit fare $f_{pt}$ is $2.75 per ride.
- Rideshare: The fare is composed of two parts: a base fare and an extra distance-based fare (Häll, 2006) defined as

---
[1] https://www.taxi-calculator.com/taxi-fare-new-york-city/259

$$f_r = f_0 + \theta d_r, \tag{17}$$

where $f_0$ is the base fare, assumed to be \$1, $\theta$ is the per-unit-distance (in kilometers) fare, assumed to be \$ $(\bar{c} + 0.1)$/km, where $\bar{c}$ is the average operating cost per kilometer traveled, taking into account driver and fuel cost. $d_r$ is the travel distance between the origin and destination of customer r. In the numerical study, distance is measured as the Euclidean distance.

- Rideshare+Transit: The fare is the total fare to use rideshare and transit, i.e., $f_{r+pt} = f_r + f_{pt}$, where $f_r$ is the fare for the rideshare leg and $f_{pt}$ is the cost of the transit leg.

We consider the morning peak-hour commuting simulation scenario from 7:00 to 9:00. Customers are randomly generated in terms of origins and destinations at the beginning of each day with a Poisson arrival intensity. We run over 20 days (iterations) to calibrate demand parameters. As the true parameters of customers' utility function are unknown, we set initial values of these parameters as $\hat{\beta}^0_{OVTT} = -0.2$, $\hat{\beta}^0_{IVTT} = -0.1$, and $\hat{\beta}^0_c = -0.1$; $(\hat{\mu}^0_{non-motorized}, \hat{\mu}^0_{auto}, \hat{\mu}^0_{pt}) = (1,1,1)$.

The simulation was implemented by Matlab on a Dell Latitude E5470 laptop with win64 OS, Intel i5-6300U CPU, 2 Cores and 8GB memory. Table 2 reports relevant reference parameter settings. The data used in this study is freely available at https://github.com/tym677.

Table 2. Reference parameter settings of the experiment

| | |
|---|---|
| Fleet size (vehicle) | 40 |
| Number of zones | 16 |
| Capacity of vehicle (customers/vehicle) | 10 |
| Idle vehicle relocation interval (minute) | 15 |
| k-nearest entry (exit) stations considered* | 4 |
| $\alpha$ (Eq.(14)) | 0.05 |

Remark: A total of $k^2 = 16$ paths are searched when determining the entry and exit stations for a rideshare+transit option generation.

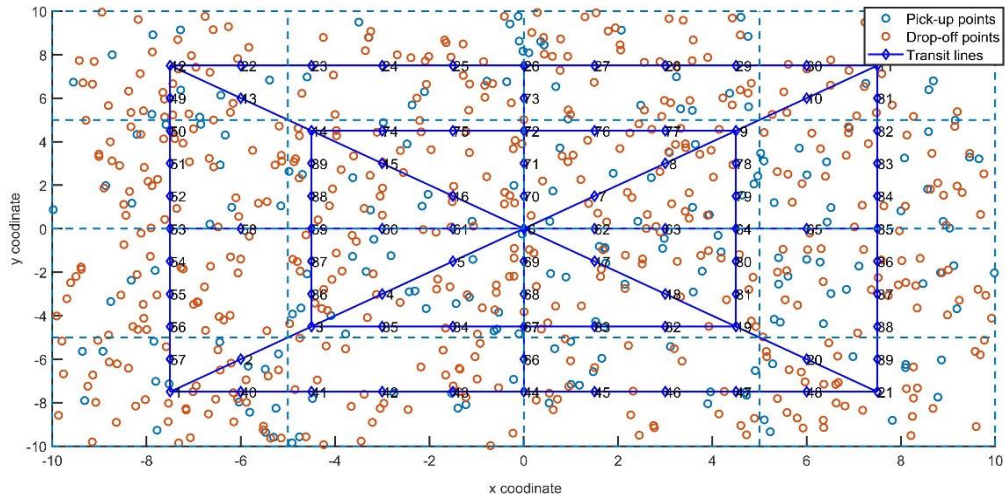

Fig. 4. Pick-up and drop-off locations of ride requests ($\lambda$=400 customers/hour)

5.2. Results

We estimate the model parameters at the end of each day using historical choice data. The maximum-likelihood approach is applied for parameter calibration. Fig. 5 reports the convergence result of the calibrated parameters of the NL model. The gap function is defined as the summation of the absolute values of the difference between the estimated and true parameters. As shown in Fig. 5, the gap falls quickly from 2.25 to 0.5 after 2 and 3 iterations (days) given different arrival intensities. The gap function stabilizes with small fluctuations from the 6$^{th}$ iteration, with a gap of around 0.25. The gap reflects a day-to-day stochastic user equilibrium fluctuation in a stochastic environment (Djavadian and Chow, 2017).

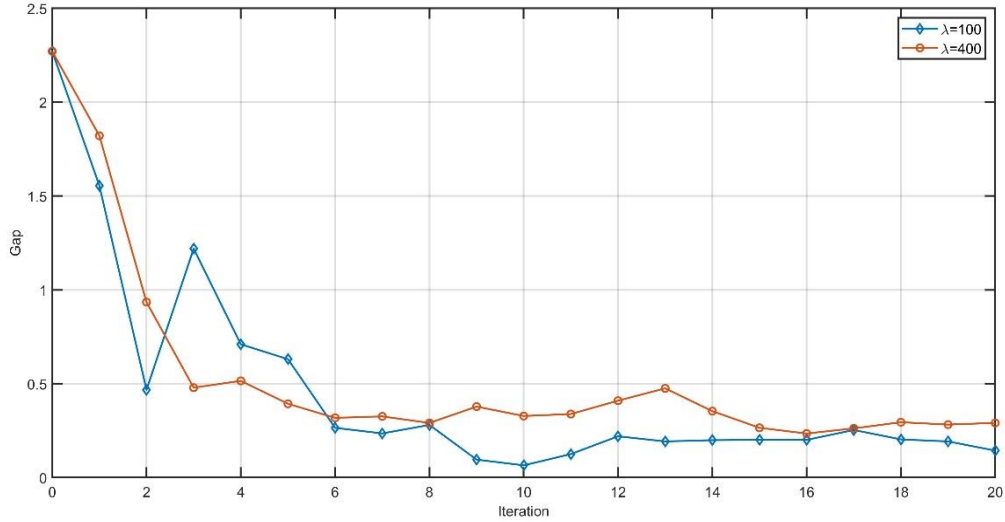

Fig. 5. Convergence of the gap function for the estimated NL model parameters.

Table 3 reports the impact of different pricing schemes on the mode shares over 'rideshare' and 'rideshare+transit'. The result shows that when customer arrival intensity is increased (i.e. $\lambda$=200 customers/hour or above), applying the dynamic pricing scheme could shift the ridership from "rideshare only" to "rideshare+transit". Due to this effect, the mean vehicle travel times (see Table 4) are reduced due to more customers being dropped off at nearby transit stations. Moreover, more vehicles would be made available due to more "rideshare+transit" users, resulting in lower waiting times to get picked up. Table 4 compares the system performances in terms of customer inconvenience and operating costs under different demand intensities. It shows that for $\lambda$=100 customers/hour, the mean waiting time and mean passenger journey time remain almost unchanged. However, the mean vehicle travel time is reduced around –5.24% due to lower ridership after applying the dynamic price scheme (i.e., –1.2% for "rideshare only" and –9.1% for "rideshare+transit", shown in Table 5).

Table 3. Mode share for systems with and without price optimization (%)

|  | $\lambda$=100 | | $\lambda$=200 | |
| --- | --- | --- | --- | --- |
|  | Rideshare | Rideshare+Transit | Rideshare | Rideshare +Transit |
| No price optimization | 22.7 | 18.5 | 21.5 | 17.0 |
| With price optimization | 22.4(-1.2%) | 16.8(-9.1%) | 20.4(-5.1%) | 17.8 (+4.7%) |
| No WTP constraints | 13.2(-41.9) | 9.9(-46.5%) | 10.1(-53.0%) | 8.1(-52.4%) |
|  | $\lambda$=400 | | $\lambda$=800 | |
|  | Rideshare | Rideshare+Transit | Rideshare | Rideshare +Transit |
| No price optimization | 18.3 | 14.1 | 11.9 | 11.1 |
| With price optimization | 17.2(-5.7%) | 15.2(+7.9%) | 11.4(-4.2%) | 11.9(+7.2%) |
| No WTP constraints | 10.9(-40.4%) | 8.8(-37.6%) | 8.4(-29.4%) | 6.4(-42.3%) |

Remark: 1.The reported values are the average mode shares over 20 iterations for a 2 hour customer arrival period. 2. WTP means willingness to pay, measured in USD.

Table 4. System performance for systems with and without price optimization

|  | $\lambda=100$ | | | $\lambda=200$ | | |
|---|---|---|---|---|---|---|
|  | WT | JT | VTL | WT | JT | VTL |
| No price optimization | 6.4 | 26.0 | 76.3 | 8.2 | 29.6 | 123.5 |
| With price optimization | 6.4 | 25.9 | 72.3 | 7.9 | 28.3 | 121.0 |
| ±% | (0%) | (-0.38%) | (-5.24%) | (-3.36%) | (-4.66%) | (-2.04%) |
|  | $\lambda=400$ | | | $\lambda=800$ | | |
|  | WT | JT | VTL | WT | JT | VTL |
| No price optimization | 13.9 | 41.2 | 172.1 | 27.3 | 59.6 | 217.7 |
| With price optimization | 12.9 | 38.6 | 168.9 | 26.2 | 56.9 | 213.2 |
| ±% | (-7.19%) | (-6.31%) | (-1.86%) | (-4.03%) | (-4.53%) | (-2.07%) |

Remark: 1. WT: mean passenger waiting time, JT: mean passenger journey time, VTL: mean vehicle travel time, defined as total vehicle travel times divided by the number of vehicles. 2. Measured in minutes.

In terms of the operator profit, Table 5 reports the impact of applying the proposed dynamic pricing method. When applying the pricing scheme, the average profit of the operator increases from +20.8% to +23.9% under different demand levels. The standard deviation of the profit is around 2.4 times higher compared to that for the system without price optimization. Fig. 6 displays the variation in the operator's profit over time. This shows that over time, the system with price optimization has significantly higher profit compared to the system without (except on day 10 for the low-demand case).

Table 5. Operator profit per iteration (day) for systems with and without price optimization

|  | $\lambda=100$ | | $\lambda=200$ | |
|---|---|---|---|---|
|  | Without price opt. | With price opt. | Without price opt. | With price opt. |
| Mean | 149.7 | 184.6 (+23.3%) | 280.3 | 347.2 (+23.9%) |
| S.D. | 10.7 | 23.9 | 12.3 | 27.6 |
| Avg. computational time per iteration (second) | 100 | 114 | 181 | 188 |
|  | $\lambda=400$ | | $\lambda=800$ | |
|  | Without price opt. | With price opt. | Without price opt. | With price opt. |
| Mean | 470.2 | 581.1 (+23.6%) | 656.3 | 793.1 (+20.8%) |
| S.D. | 14.1 | 35.0 | 21.8 | 39.3 |
| Avg. computational time per iteration (second) | 348 | 354 | 951 | 1006 |

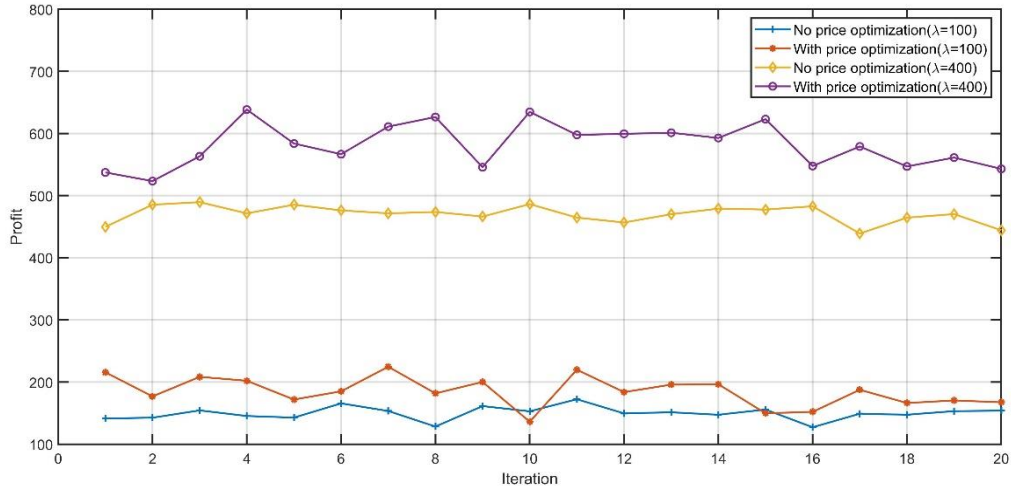

Fig. 6. Operator profit for systems with and without price optimization under different customer arrival intensities

Fig. 7 shows the impact of the proposed chance-constrained assortment pricing model on ticket prices for "rideshare only" and "rideshare+transit". The left panel in Fig. 7 shows that applying pricing optimization increases the fares of "rideshare only". However, the distribution of ticket prices for "rideshare+transit" is over a wider range and with lower prices in general compared with the initial fare without price optimization (right panel of Fig. 7). The range of ticket prices after the constrained price optimization is under 8 dollars. Note that for short-distance trips, the proposed ticket prices are relatively low due to lower customer willingness to pay for that travel distance. Furthermore, neglecting customer willingness to pay constraints would lead to unrealistically high ticket prices (around 23 dollars/ride for "rideshare only") and reduce ridership significantly under different demand levels (e.g., –40.4% (rideshare only) and –36.5% (rideshare+transit) for $\lambda = 400$ customers/hour; see Table 3). Clearly, this shows the importance of making use of customer willingness to pay as a constraint to frame the ticket price within acceptable values for customers, in order to keep the proposed MOD services competitive in a multimodal transport market.

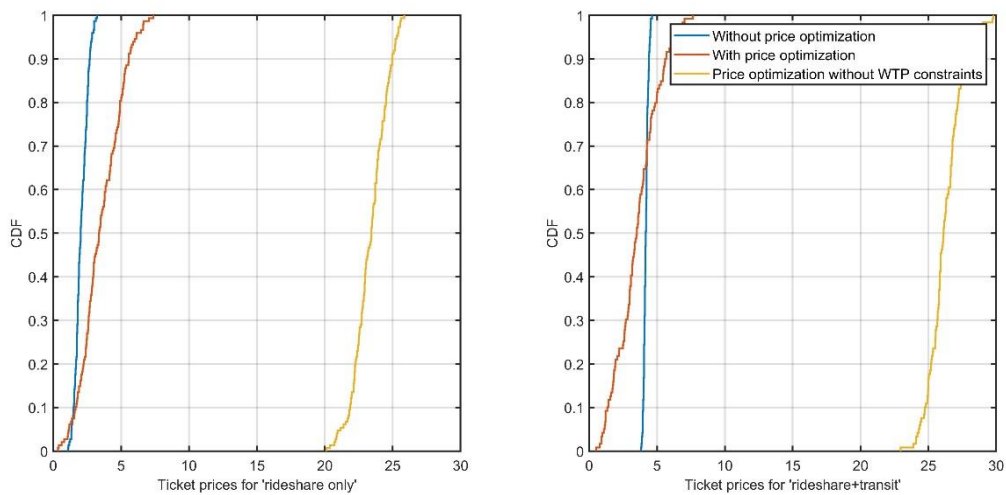

Fig. 7. Cumulative probability functions of ticket prices for systems with and without price optimization ($\lambda = 400$ customers/hour, WTP means willingness-to-pay)

Fig. 8 depicts the mode shares for different transport modes in the high demand case ($\lambda$=400 customers/hour). We find that the proposed pricing model would shift "rideshare only" users to "rideshare+transit" while keeping the mode shares of the other modes (almost) unchanged. However, the classical unconstrained assortment pricing optimization would result in a significant mode shift from the rideshare options to the other modes (i.e., car, taxi, and bike) due to its unrealistically high ticket prices.

Clearly, the proposed pricing approach could significantly increase the operator's profit while reducing user inconvenience and operating costs by setting up a more attractive price for "rideshare+transit" compared to "rideshare only".

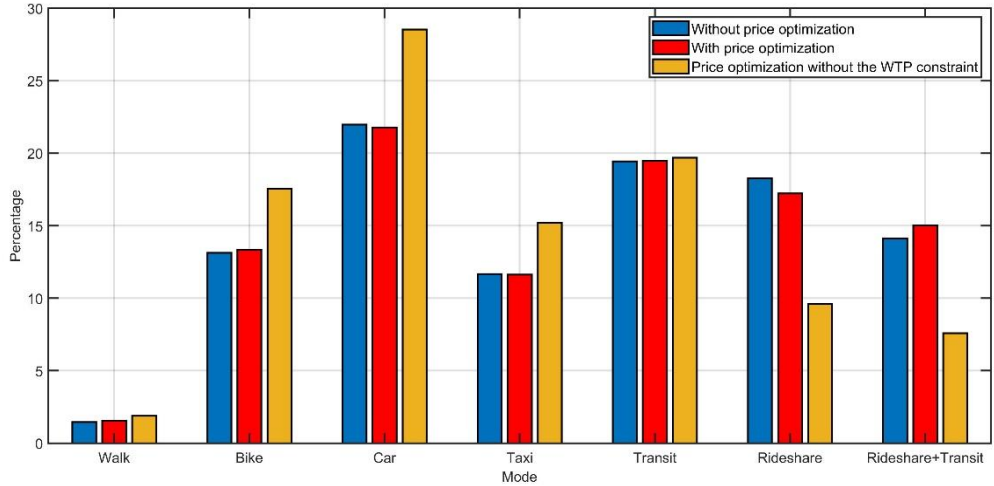

Fig. 8. Mode share for systems with and without price optimization ($\lambda$=400)

5.3. Sensitivity analysis

In this section, we investigate how different system parameters, i.e., vehicle capacity and the variance of customer willingness to pay, impact the performance of the proposed model. We analyze the impacts for two different customer arrival intensities corresponding to low ($\lambda = 100$ customers/hour) and high ($\lambda = 400$ customers/hour) demand levels under the same simulation settings as described in Section 5.1.

5.3.1. Influence of vehicle capacity

We vary the capacity of vehicles over three scenarios: 5, 10, and 15 passengers/vehicle. Table 6 compares the average profit and performance of systems with and without price optimization. Overall, the results confirm the benefit of the proposed pricing model on operator profit (i.e., +23.3% to +24.9%) given different capacities of vehicles.

Table 6. The influence of vehicle capacity on operator profit and system performance with and without price optimization

|  |  | $q_v = 5$ | | | | $q_v = 10$ | | | | $q_v = 15$ | | | |
|---|---|---|---|---|---|---|---|---|---|---|---|---|---|
|  | P.O. | Profit | WT | JT | VTL | Profit | WT | JT | VTL | Profit | WT | JT | VTL |
| | Without | 149.7 | 6.4 | 26.0 | 76.3 | 149.7 | 6.4 | 26.0 | 76.3 | 149.7 | 6.4 | 26.0 | 76.3 |
| $\lambda$=100 | With | 184.6 | 6.4 | 25.9 | 72.3 | 184.6 | 6.4 | 25.9 | 72.3 | 184.6 | 6.4 | 25.9 | 72.3 |
| | ±% | +23.3 | -1.5 | -0.5 | -5.3 | +23.3 | -1.5 | -0.5 | -5.3 | +23.3 | -1.5 | -0.5 | -5.3 |

|  | | Without | 469.8 | 14.0 | 41.3 | 172.4 | 470.2 | 13.9 | 41.2 | 172.1 | 468.1 | 13.8 | 40.8 | 170.7 |
|---|---|---|---|---|---|---|---|---|---|---|---|---|---|---|
| $\lambda =400$ | | With | 583.0 | 13.3 | 39.2 | 169.3 | 581.1 | 13.0 | 38.8 | 168.6 | 583.0 | 13.3 | 39.2 | 169.3 |
| | | ±% | +24.1 | -4.4 | -4.9 | -1.8 | +24.9 | -6.9 | -5.9 | -2.1 | +24.5 | -3.7 | -4.0 | -0.8 |

Remark: 1. P.O: price optimization. 2. WT: mean passenger waiting time, JT: mean passenger journey time, VTL: mean vehicle travel time.

5.3.2. Influence of $\sigma$

Let us look at how the variance of customer willingness to pay influences the performance of the proposed pricing model. We vary $\sigma$ over three values: 0, 1, and 4. Table 7 shows that increasing $\sigma$ results in a significantly higher average profit for the operator. For ticket prices, we find that the higher the variance of customer willingness to pay, the higher the average ticket prices proposed by the dynamic pricing scheme. The price increases from around $3.5 to $6.8 if $\sigma$ increases from $0 to $4. However, for mode shares of the MOD service, the price decreases accordingly from –6.5% to –7.7% for $\lambda$=100 and $\lambda$=400 (customers/hour), respectively. Fig. 9 shows the day-to-day variation in operator profit based on different values of $\sigma$. Again, $\sigma$ needs to be calibrated in realistic applications to capture the variation of customer's willingness to pay to achieve the desired objective.

Table 7. The influence of $\sigma$ on the operator's average profit per iteration (day) with pricing optimization

|  | $\lambda$=100 | | | $\lambda$=400 | | |
|---|---|---|---|---|---|---|
|  | $\sigma$=0 | $\sigma$=1 | $\sigma$=4 | $\sigma$=0 | $\sigma$=1 | $\sigma$=4 |
| Mean | 184.6 | 304.0 | 408.5 | 581.1 | 975.0 | 1329.7 |
| S.D. | 20.2 | 30.4 | 34.3 | 31.5 | 39.4 | 56.0 |

Table 8. The influence of $\sigma$ on the profit, ticket prices, and mode share of the MOD services per iteration

|  |  | $\lambda$=100 | | $\lambda$=400 | |
|---|---|---|---|---|---|
|  |  | Mean | S.D | Mean | S.D |
| $\sigma$=0 | Profit | 184.6 | 20.2 | 581.1 | 31.5 |
|  | Price | 3.52 | 1.74 | 3.55 | 1.54 |
|  | Mode share (%) | 39.2 | 2.8 | 32.4 | 1.3 |
| $\sigma$=1 | Profit | 304 | 30.4 | 975 | 39.4 |
|  | Price | 5.17 | 1.75 | 5.21 | 1.51 |
|  | Mode share (%) | 38.8 | 3.0 | 31.3 | 1.5 |
| $\sigma$=4 | Profit | 408.5 | 34.3 | 1329.7 | 56 |
|  | Price | 6.86 | 1.79 | 6.84 | 1.56 |
|  | Mode share (%) | 36.7 | 2.9 | 29.9 | 1.1 |

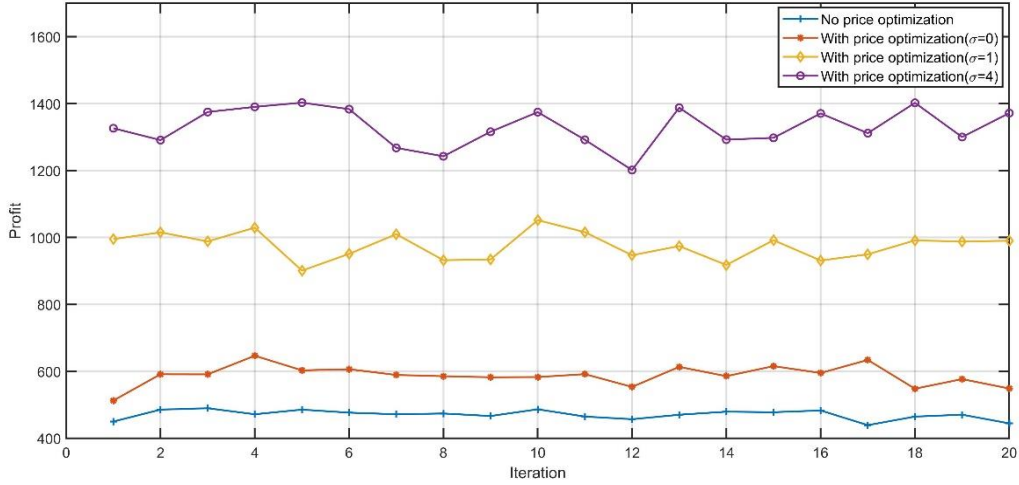

Fig. 9. Comparison of operator profit under different $\sigma$ ($\lambda = 400$)

## 6. Conclusions

In this study, we propose a dynamic integrated microtransit system with customer demand learning and dynamic pricing based on user's willingness-to-pay constraints. We consider two kinds of MOD services in a multimodal transport market: classical door-to-door rideshare and an integrated rideshare with transit transfer service. In the proposed dynamic pricing model, customer choice behavior is integrated for constrained assortment pricing optimization to maximize the expected profit of operators. Under the assumption of the ability to collect customer choice intentions from the operator's service platform/application, we demonstrate that the choice model can be calibrated over time. However, when only purchased MOD service mode choices are observable, one can apply the methodology of Talluri and Van Ryzin (2006, Chapter 9) to calibrate the choice model based on the incomplete data.

We demonstrate the proposed methodology through a numerical example to quantify the effect of dynamic pricing. While we did not model the endogenous network congestion effect due to day-to-day traffic flow changes on the multimodal network, this paper did consider the interactions of supply and demand in a multimodal market by integrating customer mode choice behavior into a constrained pricing scheme. We tested the proposed model under different scenarios in terms of customer arrival intensity, vehicle capacity, and the variance of user willingness to pay. Results suggest that the proposed chance-constrained assortment price optimization model allows increasing operator profit while keeping the proposed ticket prices acceptable. However, a classical non-constrained assortment price optimization approach would result in unrealistically high ticket prices and a significant reduction in the ridership of MOD services. Results show that the impact of the proposed pricing model could be to reduce system operating costs and customer inconvenience due to a more attractive ticket price setting for the "rideshare+transit" option after price optimization.

In terms of policy implication, recent studies shows the importance of pricing context and customer's behavior in response to the trade-off of ticket price and level of service for MOD service planning (Chow et al., 2020). Future research could improve this work by considering an agent-based stochastic user equilibrium and more realistic modeling of transportation system dynamics. From the government perspective, it is interesting to maximize social welfare based on the dynamic ticket prices setting. Other online survey methodologies (e.g., smartphone survey) can be designed to collect customers' willingness to pay for different user groups. For mode choice behavior modeling, it is also desirable to incorporate an individual's socio-demographic variables to explain the heterogeneity of travel preference, among many other practical aspects, to improve the realistic application of the proposed modeling approach.